\documentclass[11pt]{article}

\usepackage{graphicx,amsmath,amsfonts,amssymb,dcolumn,amsthm,slashed,bbm,xspace,pgffor,tikz,mathbbol,latexsym,enumerate}
\usepackage{soul}
\usepackage[colorlinks,citecolor=red,urlcolor=cyan,linkcolor=blue]{hyperref}
\usepackage[utf8]{inputenc}
\usepackage[english]{babel}
\usepackage{lmodern}
\usepackage{microtype}
\usepackage{cite}
\usepackage[normalem]{ulem}

\numberwithin{equation}{section}

\begin{document}

\title{\textbf{Renormalizability of a first principles Yang-Mills center-vortex ensemble
}}

\author{\textbf{D.~Fiorentini}$^1$\thanks{diego\_fiorentini@id.uff.br},\ \textbf{D.~R.~Junior}$^{1,2}$\thanks{davidjunior@id.uff.br},\ \textbf{L.~E.~Oxman}$^1$\thanks{leoxman@id.uff.br},\  \textbf{R.~F.~Sobreiro}$^1$\thanks{rodrigo\_sobreiro@id.uff.br}\\\\
\textit{{\small$^1$ Universidade Federal Fluminense, Instituto de F\'isica,}}\\
\textit{{\small Av. Litorânea s/n, 24210-346, Niter\'oi, RJ, Brasil.}}\\
\textit{{\small $^2$ Universit\"at T\"ubingen, Institut f\"ur Theoretische Physik}}\\
\textit{{\small  Auf der Morgenstelle 14, 72076, Tübingen, Deutschland.}}}

\date{}
\maketitle

\begin{abstract}
Recently, a new procedure to quantize the $SU(N)$ Yang-Mills theory in the nonperturbative regime was proposed. The idea is to divide the configuration space $\{A_\mu\}$ into sectors labeled by different topological degrees of freedom and fix the gauge separately on each one of them. As Singer's theorem  on gauge copies only refers to gauge fixing conditions that are global in $\{A_\mu\}$, this construction might avoid the Gribov problem. In this work, we present a proof of the renormalizability in the center-vortex sectors, thus establishing the calculability of the Yang-Mills center-vortex ensemble.
\end{abstract}

\newpage

\tableofcontents

\newpage

\section{Introduction}

In 1978, I.~M.~Singer showed that any gauge-fixing condition in $SU(N)$ Yang-Mills (YM) theory that is global in configuration space $\{A_\mu\}$ will necessarily contain Gribov copies  \cite{Singer:1978dk,Gribov:1977wm,Sobreiro:2005ec}. This is the fundamental reason behind  the infrared problems faced when trying a quantization in the continuum. 
In the last many years, the main approach to circumvent this problem has been based on the restriction of the path integral to the first Gribov region, the ensuing Gribov-Zwanziger quantization procedure  \cite{Zwanziger:1989mf,Zwanziger:1992qr,Dudal:2005na}, as well as its refinement and improvement \cite{Dudal:2008sp,Dudal:2011gd,Capri:2015ixa,Capri:2016aqq}. It is interesting to note that, in his work, Singer pointed to a different procedure based on a locally finite  open covering $\{\vartheta_\alpha\}$ of  the total space of gauge field configurations  $\{A_\mu\}$, namely,
\begin{equation}
     \{A_\mu\}=\cup_{\alpha} \vartheta_\alpha\;,
\end{equation}
together with a subordinate partition of unity \cite{rudin,koba}
\begin{gather}
    \sum_\alpha \rho_{_\alpha}(A_\mu) = 1 \makebox[.5in]{,} \forall A_\mu \in \{A_\mu\} \;,
\end{gather}
where the support of the function $\rho_\alpha$ is $\vartheta_\alpha$. Introducing this identity, the YM partition function can be rewritten as 
\begin{align}
    Z_{{\rm YM}}=\sum_\alpha Z_{(\alpha)}\makebox[.5in]{,}Z_{(\alpha)}=\int_{\vartheta_\alpha} [DA]\,\rho_{\alpha}(A)\,e^{-S_{{\rm YM}}[]A]}\;,
\end{align}
 where $S_{{\rm YM}}$ is the YM action, see Eq. \eqref{ym1}. Note that, in each term, the path-integral can be done on the  support of $\rho_\alpha (A) $. 
Now, by choosing the components of the covering $\vartheta_\alpha$ such that they admit local cross sections \begin{gather}
    f_\alpha(A)=0 \;,
\end{gather}
without copies, the usual Faddeev-Popov procedure can be safely implemented on each term $Z_{(\alpha)}$
\begin{gather}
    Z_{\rm YM} = \sum_\alpha \int_{\vartheta_\alpha} [DA]\,\rho_\alpha(A)\,e^{-S_{\rm YM}}\,\delta(f_\alpha(A))\,{\rm Det}\frac{\delta f_\alpha(A^U)}{\delta U}\Bigg|_{U=I}\,\;.\label{general}
\end{gather}Over the years, this possibility was overlooked, certainly because of the difficulties to identify and characterize this covering and effectively implement the partition of unity. Along this line, if the covering were a partition of $\{A_\mu\}$,
\begin{gather}
    \{A_\mu\} = \cup_\alpha \vartheta_\alpha \makebox[.5in]{,} \vartheta_\alpha \cap \vartheta_\beta = \emptyset\makebox[.3in]{,}  \alpha\neq\beta \;, \label{partition}
\end{gather}
then $\rho_\alpha(A)$ would be a characteristic function $\theta_{\alpha}(A)$, which is one if $A_\mu\in\vartheta_\alpha$ and is zero otherwise. This case was precisely implemented in Ref. \cite{Oxman:2015ira}, by introducing a partition of $\{A_\mu\}$ induced by an equivalence relation. The main idea is to correlate the gauge field $A_\mu$ with auxiliary adjoint scalar fields $\zeta_I(A)$ by means of a classical equation of motion.  Following a generalized polar decomposition, the tuple $\zeta_I(A)$ was written in terms of a modulus $q_I(A)$ and a phase mapping $S(A)\in SU(N)$. These mappings can be divided into equivalence classes
induced by the equivalence relation  $S' \sim S $, if $S'=US$ for some regular $U \in SU(N)$. As the phases $S(A)$ are not always regular, this leads to a nontrivial partition $\vartheta_{S_0} =[S_0]$. The labels $S_0$ are the different class representatives, which correspond to the different distributions of topological defects such as oriented and nonoriented center vortices. Then, the gauge is separately fixed with the sector-dependent condition 

\begin{gather}
  f_{S_0}(A)=[S_0^{-1}\zeta_I(A) S_0,T_I]=0\;, \label{gaugecond} 
\end{gather}
which ensures that the auxiliary fields have the form $\zeta_I(A)  = S_0 q_I (A) S_0^{-1}$, where the tuple $q_I(A)$ is a pure modulus. The conditions for this procedure to be well-defined were extensively discussed in Ref. \cite{PhysRevD.103.114010}. The YM partition function is then given by Eq. \eqref{general}, in the form,
\begin{gather}
    Z_{\rm YM} = \sum_{S_0} Z^{S_0}_{\rm YM} \makebox[.5in]{,} Z^{S_0}_{\rm YM} =  \int_{\vartheta_{S_0}} [DA] \,e^{-S_{\rm YM}}\,
    \delta(f_{S_0} (A))\,{\rm Det}\frac{\delta f_{S_0}(A^U)}{\delta U}\Bigg|_{U=I}\,\;.\label{general-1}
\end{gather} 
Averages of observables $O[A]$ may be similarly evaluated as the sum of partial contributions:
\begin{align}
    \langle O\rangle = \sum_{S_0}\frac{Z^{S_0}_{\rm YM}}{Z_{\rm YM}}\langle O\rangle_{S_0}\makebox[.5in]{,}\langle O\rangle_{S_0}= \int_{\vartheta_{S_0}} [D A_\mu]\, e^{-S_{\rm YM}}O[A]\delta(f_{S_0}(A))\,{\rm Det}\frac{\delta f_{S_0}(A^U)}{\delta U}\Bigg|_{U=I}\label{ymensemble}\;.
\end{align}
These formulas show how a proper nonperturbative definition of the YM path-integral, performed as a superposition of local contributions, could lead to an ensemble. Indeed, center-vortex ensembles are known to be  relevant for describing the confining properties of YM theory in the infrared regime \cite{DelDebbio:1996lih,Langfeld:1997jx,DelDebbio:1998luz,Faber:1997rp,deForcrand:1999our,Ambjorn:1999ym,Engelhardt:1999fd,Engelhardt:1999xw,Bertle:2001xd,Reinhardt:2001kf,Gattnar:2004gx}. A more concrete connection would be established by computing an approximation to the partial contributions $Z^{S_0}_{\rm YM}$. Of course, to make sense of the formal expressions, a fundamental requirement is renormalizability. In Ref. \cite{Fiorentini:2020tcn}, relying on the algebraic method, we showed that this property is valid in the vortex-free sector ($S_0=I$), where the possible counter-terms are restricted by the Ward Identities of the gauge-fixed action. In this work, we present a proof of renormalizability for the center-vortex sectors. 

In section 2, we present some preliminary definitions and the construction of the gauge-fixed action in the vortex-free sector from a BRST perspective.  In section 3, we extend this procedure to a general sector labeled by center vortices, and we introduce the required boundary conditions in a way that maximizes the symmetries of the full action. In section 4, we list these symmetries and use them to establish the renormalizability of a general center-vortex sector $\vartheta_{S_0}$. Finally, in section 5 we present our conclusions.

\section{Preliminary definitions and the vortex-free sector}

To construct the complete action, in a BRST formal manner \cite{Piguet:1995er}, we start by defining the usual YM action in 4-dimensional Euclidean spacetime\footnote{In this paper, we employ a condensed notation for integrals as $\int_x\equiv\int d^4x$.}
\begin{equation}
    S_{YM}=\frac{1}{4}\int_xF^a_{\mu\nu}F^a_{\mu\nu}\;,\label{ym1}
\end{equation}
where $F^a_{\mu\nu}=\partial_\mu A^a_\nu-\partial_\nu A^a_\mu+gf^{abc}A_\mu^bA_\nu^c$ is the field strength for the $SU(N)$ gauge-field $A^a_\mu$, while $g$ is the coupling constant. Lower case greek and latin indices take the values $\{0,1,2,3\}$ and $\{1,\ldots,N^2-1\}$, respectively.

Another field naturally appearing in YM theory is the Faddeev-Popov ghost field $c^a$. This field appears in the gauge-fixing procedure and, together with $A_\mu$, can be interpreted geometrically \cite{Daniel:1979ez,Nakahara:2003nw,Bertlmann:1996xk}. For now, we can define the nilpotent BRST operator $s$ and the corresponding BRST transformations of the fields
\begin{eqnarray}
    sA^a_\mu&=&\frac{i}{g}D_\mu^{ab}c^c\;,\nonumber\\
    sc^a&=&-\frac{i}{g}f^{abc}c^bc^c\;,\label{brst1}
\end{eqnarray}
with the covariant derivative defined as $D^{ab}_\mu=\delta^{ab}\partial_\mu-gf^{abc}A_\mu^c$.

To fix the gauge in the vortex-free sector, we follow the procedure developed in \cite{Oxman:2015ira,Fiorentini:2020tcn}. The first step is to introduce a set of auxiliary fields $\zeta_I^a,b_I^a,c^a_I,\bar{c}^a_I$, where the flavor index takes values in $\{1,2,\ldots,N^2-1\}$. In order to keep the physical degrees of freedom of pure YM theory unaltered, these fields are introduced as BRST doublets \cite{Piguet:1995er},
\begin{eqnarray}
    s\zeta^a_I&=&if^{abc}\zeta^b_Ic^c+c^a_I\;,\nonumber\\
    sc^a_I&=&-if^{abc}c^b_Ic^c\;,\nonumber\\
    s\bar{c}^a_I&=&-if^{abc}\bar{c}^b_Ic^c-b^a_I\;,\nonumber\\
    sb^a_I&=&if^{abc}b^b_Ic^c\;.\label{brst2}
\end{eqnarray}
Moreover, a set of BRST doublet parameters $\{\mu,U,\kappa,\mathcal{K},\lambda,\Lambda\}$ are introduced
\begin{eqnarray}
s\mu^2&=&U^2\;,\nonumber\\ sU^2&=&0\;,\nonumber\\
s\kappa &=& \mathcal{K}\;,\nonumber\\ s\mathcal{K}&=&0\;,\nonumber\\
s\lambda &=& \Lambda\;,\nonumber\\
s\Lambda &=&0\;.\label{brst3}
\end{eqnarray}
 The parameters $\mu,\kappa, $ and $\lambda$ are required to implement the $SU(N)\to Z(N)$ spontaneous symmetry breaking in the auxiliary action, thus producing the correlation between $A_\mu$ and the phases $S(A)$ containing defects. Their respective doublet partners $U^2,\mathcal{K},\Lambda$ are required to guarantee that the observables remain independent from the gauge-fixing parameters \cite{Piguet:1984js,Piguet:1995er}. The auxiliary action is introduced as a BRST-exact term in the form
\begin{eqnarray}
    S_{\rm aux}&=&-s\int_x\left(D^{ab}_\mu\bar{c}^b_ID_\mu^{ac}\zeta^c_I+\mu^2\bar{c}^a_I\zeta^a_I+\kappa f^{abc}f_{IJK}\bar{c}^a_I\zeta^b_J\zeta^c_K+\lambda\gamma^{abcd}_{IJKL}b^a_I\zeta^b_J\zeta^c_K\zeta^d_L\right)\nonumber\\
    &=&\int_x\left( D_\mu^{ab}b_I^b D_\mu^{ac} \zeta_I^c+ D_\mu^{ab}\bar{c}_I^b D_\mu^{ac} c_I^c+\mu^2 \left(\bar{c}^a_I c^a_I + b^a_I \zeta^b_I\right) + \kappa f_{IJK}f^{abc}( b_I^a \zeta_J^b \zeta^c_K -2\bar{c}_I^a \zeta^b_K c^c_J)+\right.\nonumber\\
    &+&\left.  \lambda\gamma_{IJKL}^{abcd}\left(b_I^a\zeta_J^b\zeta_K^c\zeta_L^d+3\bar{c}_I^ac_J^b\zeta_K^c\zeta_L^d\right)- U^2 \bar{c}^a_I \zeta^a_I-\Lambda f^{abc}f^{cde}\bar{c}^a_I \zeta^b_J \zeta^d_I \zeta^e_J  -\mathcal{K}f^{IJK} f^{abc} \bar{c}^a_I \zeta^b_J \zeta^c_K\right)\;.\nonumber\\
    \label{aux1}
\end{eqnarray} 
Here, $\gamma$ is a general color-flavor tensor, which is invariant under the adjoint global symmetry group ${\rm Ad}(SU(N))$. For later use, when implementing the different symmetries, we shall consider  tensors formed by antisymmetric structure constants and Kronecker deltas such that $s\gamma^{abcd}_{IJKL}b_I^a\zeta_J^b\zeta_K^c\zeta_L^d$ is $c$-independent, i.e. \begin{eqnarray}
    \gamma^{mbcd}_{IJKL}f^{ame}+\gamma^{amcd}_{IJKL}f^{bme}+\gamma^{abmd}_{IJKL}f^{cme}+\gamma^{abcm}_{IJKL}f^{dme}=0\;.
\end{eqnarray}
The gauge fixing \emph{per se} is performed in an indirect way by imposing some condition on the auxiliary fields (see for instance \cite{Oxman:2015ira,Fiorentini:2020tcn}). With this purpose, the usual BRST doublet $\{\bar{c}^a,b^a\}$ is introduced \cite{Piguet:1995er}
\begin{eqnarray}
    s\bar{c}^a&=&-b^a\;,\nonumber\\
    sb^a&=&0\;,\label{brst4}
\end{eqnarray}
with $\bar{c}^a$ being the Faddeev-Popov anti-ghost field and $b^a$ the Lautrup-Nakanishi field. In the vortex-free sector, the representative $S_0$ in Eq. \eqref{gaugecond} can be taken as the identity, so the gauge fixing condition reads
\begin{equation}
  f^{abc}\eta^b_I\zeta^c_I=0\makebox[.5in]{,} \eta_I=vT_I\;.\label{gf1}
\end{equation}
The parameter $v$ has mass dimension. Indeed, the field $\eta_I$ can be thought of as a reference element in the classical vacua manifold $\mathcal{M}$ of the auxiliary action\cite{Oxman:2015ira,Fiorentini:2020tcn,PhysRevD.103.114010}. The gauge fixing is essentially a condition setting the local frame $\zeta^a_I T^a \in \mathfrak{su}(N)$ to lie as close as possible to the global frame $v T_I$. Such a condition is realized by the gauge-fixing action
\begin{eqnarray}
    S_{\rm gf}&=&-s\int_xif^{abc}\bar{c}^a\eta^b_I\zeta^c_I\;,\nonumber\\
    &=&\int_x\left[if^{abc} \left(b^a\eta_I^b\zeta_I^c +\bar{c}^a\eta_I^bc_I^c \right)+f^{ecd}f^{eba}\bar{c}^a\eta_I^b\zeta_I^cc^d\right]\;.\label{gf2}
\end{eqnarray}
The full gauge-fixed action in the vortex-free sector then reads
\begin{equation}
    S_{\rm vf}=S_{\rm YM}+S_{\rm aux}+S_{\rm gf}\;.\label{ym2}
\end{equation}
Note that, as the terms $S_{\rm aux}$ and $S_{\rm gf}$ are BRST exact, the theory continues to be pure YM, in spite of the SSB properties of the auxiliary sector. Another important feature of action \eqref{ym2} is a global flavor symmetry, which implies an extra conserved  charge (besides the ghost number), the $\mathcal{Q}$-charge. Such symmetry and others play a crucial role in the proof of renormalizability of the vortex-free sector \cite{Fiorentini:2020tcn}. For completeness and further use, we display in Tables \ref{table1} and \ref{table2} the quantum numbers of fields and parameters so far introduced.

\begin{table}[h]
\centering
\begin{tabular}{|c|c|c|c|c|c|c|c|c|c|c|c|c|c}
\hline
Fields &$A_\mu$&$\zeta_I$&$c_I$&$\bar{c}_I$&$b_I$&$\eta_I$&$\bar{c}$&$c$&$b$&$\xi_I$&$\lambda_I$\\
\hline\hline
Mass dimension &1&1&1&1&1&1&2&0&2&1&1\\
\hline
Ghost number&0&0&1&$-1$&0&0&$-1$&1&0&-1&0\\
\hline
$\mathcal{Q}$-charge &0&$1$&1&$-1$&$-1$&$-1$&$0$&$0$&$0$&-1&$-1$\\
\hline
Nature &B&B&F&F&B&B&F&F&B&F&B\\
\hline
\end{tabular}
\caption{Quantum numbers of the fields.}
\label{table1}
\end{table}

\begin{table}[ht]
\centering
\begin{tabular}{|c|c|c|c|c|c|c|}
\hline
Parameters &$U^2$&$\mathcal{K}$&$\Lambda$&$\mu^2$&$\kappa$&$\lambda$\\
\hline\hline
Mass dimension &2&1&0&2&1&0\\
\hline
Ghost number &1&1&1&0&0&0\\
\hline
$\mathcal{Q}$-charge &0&$-1$&$-2$&$0$&$-1$&$-2$\\
\hline
Nature &F&F&F&B&B&B\\
\hline
\end{tabular}
\caption{Quantum numbers of the parameters.}
\label{table2}
\end{table}

\section{Center-vortex sectors}

Let us consider a sector labeled by $n$ elementary center vortices located at arbitrary closed surfaces $\Omega_i$. When they carry the same fundamental weight $\beta$, the associated phase can be written as $S_0=e^{i\chi\beta\cdot T} $, where $\chi$ is multivalued when going around $\Omega = \Omega_1\cup\dots\cup\Omega_n$. Here, we use the definition $\beta\cdot T \equiv \beta_q T_q$, where $\beta_q$ are the components of the $(N-1)$-tuple $\beta$, and $T_q$ are the Cartan subalgebra generators\footnote{See Appendix \ref{Ap1} where we describe the conventions used for the Lie algebra generators}. In this case, because of the sector-dependent gauge condition in Eq. \eqref{gaugecond}, the typical gauge-fixed configurations of auxiliary fields will be of the form $\zeta_I=S_0q_I S_0^{-1}$, with $[q_I,T_I]=0$. Then, to assure regularity, the components of $\zeta_I$ that rotate under $S_0$ must vanish at $\Omega$. These are given by the fields $\zeta_I^{\alpha}$ and $\zeta_I^{\bar{\alpha}}$ along the off-diagonal directions $T_{\alpha}, T_{\bar{\alpha}}$, with $\alpha\cdot\beta \neq 0$. The $(N-1)$-tuples $\alpha$ are the roots of $\mathfrak{su}(N)$, i.e. they are formed by eigenvalues of the adjoint action of the Cartan generators (see Appendix \ref{Ap1}).  A well-known manner to implement this type of boundary condition is to introduce a $\delta-$functional in the partition function, and exponentiate it using auxiliary fields that only exist in $\Omega$ \cite{Golestanian:1998bx},
\begin{eqnarray}
\prod\limits_{\gamma}\delta_\Omega(\zeta_I^{\gamma})\delta_\Omega(\zeta_I^{\bar{\gamma}}) = \int [D\lambda] \; e^{i\sum\limits_{\gamma}\int d\sigma_1d\sigma_2 \sqrt{g(\sigma_1,\sigma_2)} \left(\lambda_I^{\gamma}(\sigma_1,\sigma_2)\zeta_I^{\gamma}(x(\sigma_1,\sigma_2))+\lambda_I^{\bar{\gamma}}(\sigma_1,\sigma_2)\zeta_I^{\bar{\gamma}}(x(\sigma_1,\sigma_2))\right)}\;,
\end{eqnarray}
where $x(\sigma_1,\sigma_2)$ is a parametrization of $\Omega$, $\lambda^{\gamma}_I$ and $\lambda^{\bar{\gamma}}_I$ are auxiliary fields, g is the determinant of the worldsheet metric, and we are denoting by $\gamma$ the roots that satisfy $\gamma\cdot\beta\neq 0$. By introducing a source localized on $\Omega$, this expression can also be written in terms of a field $\lambda_I$ defined on the whole spacetime
\begin{equation}
\prod\limits_{\gamma}\delta_\Omega(\zeta_I^{\gamma})\delta_\Omega(\zeta_I^{\bar{\gamma}})=\int [D\lambda]\; e^{i\int dx\;J_{\Omega}(x)\sum\limits_{\gamma}\left(\lambda_I^{\gamma}(x) \zeta_I^{\gamma}(x)+\lambda_I^{\bar{\gamma}}(x) \zeta_I^{\bar{\gamma}}(x)\right)}\;, \label{orig}
\end{equation}
\begin{equation}
J_{\Omega}(x)=\int d\sigma_1d\sigma_2\sqrt{g(\sigma_1,\sigma_2)}\delta(x-x(\sigma_1,\sigma_2))\;.
\end{equation}
This procedure was proposed in Ref. \cite{Fiorentini:2020tcn}. However, these terms break the color-flavor symmetry, which would allow too many new counter-terms in the renormalizability analysis. For instance, the single term $\bar{c}^ac^a$ would generate $(N^2-1)^2$ independent contributions. To circumvent this problem we can invoke the Symanzik method \cite{Symanzik:1969ek} to promote $J_{\Omega}$ to a set of generic Schwinger sources $J^a(x)$, so the color-flavor symmetry can be restored. At the end, we choose $J^a$ so as to recover the initial theory. One possibility to perform the trick is to consider the replacement
\begin{eqnarray}
&\prod\limits_{\gamma}\delta_\Omega(\zeta_I^{\gamma})\delta_\Omega(\zeta_I^{\bar{\gamma}})\rightarrow \int [D\lambda]\; e^{-\int dx\;f^{abc}J^a\lambda_I^b\zeta_I^c}\;.\label{orig1}
\end{eqnarray}
Expression \eqref{orig} is then recovered by setting the source $J^a$ to its physical values, namely,
\begin{eqnarray}
    J^{\alpha}\Big|_{\rm phys}&=&J^{\bar{\alpha}}\Big|_{\rm phys}\;=\;0\;,\nonumber\\ J^q\Big|_{\rm phys}&=&i\beta_q\int d\sigma_1d\sigma_2\sqrt{g(\sigma_1,\sigma_2)}\delta(x-x(\sigma_1,\sigma_2))\;.\label{physource}
\end{eqnarray}
In this case, we have
\begin{equation}
    f^{abc}J^a\lambda_I^b\zeta_I^c=J^qf^{qbc}\lambda_I^b\zeta_I^c\label{exp}\;,
\end{equation}
and taking into account that the only structure constants which contribute are $f^{q\alpha\bar{\alpha}}=\alpha|_q $, Eq. \eqref{exp} becomes
\begin{eqnarray}
\sum_{\alpha>0}J\cdot\alpha (\lambda_I^{\alpha}\zeta_I^{\bar{\alpha}}-\lambda_I^{\bar{\alpha}}\zeta_I^{\alpha})=\sum_{i=1}^n\sum_{\alpha>0} i\beta\cdot\alpha\int d\sigma_1d\sigma_2\sqrt{g(\sigma_1,\sigma_2)}\delta(x-x(\sigma_1,\sigma_2))(\lambda_I^{\alpha}\zeta_I^{\bar{\alpha}}-\lambda_I^{\bar{\alpha}}\zeta_I^{\alpha})\;.
\end{eqnarray}
Since the scalar product $\alpha\cdot\beta$ is either $1,-1,$ or $0$, the desired expression is recovered. However, we still need to worry about the BRST invariance of these boundary conditions, which imply $s\zeta_I^\gamma=s\zeta_I^{\bar{\gamma}}=0$ on the vortex surface \cite{Moss:1996ip}.
For this purpose, it is convenient to work with BRST doublets and write the $J^a$-term as a BRST-exact quantity. Thus, we introduce the auxiliary field  $\xi_I$ such that the pair $\{\lambda_I,\xi_I\}$ forms a BRST doublet 
\begin{eqnarray}
    s\xi_I^a&=&\lambda_I^a\;,\nonumber\\ s\lambda_I^a&=&0\;.\label{brst5}
\end{eqnarray}
This ensures that these fields cannot be part of the physical spectrum of the theory \cite{Piguet:1995er}. The source $J^a$ is assumed to be BRST invariant. Hence,
\begin{equation}
   S_J=s\int_x f^{abc}J^a\xi_I^b\zeta_I^c=\int_x f^{abc}J^a\left[\lambda_I^b\zeta_I^c-\xi^b_I\left(if^{cde}\zeta^c_Ic^e+c^c_I\right)\right]\;.
\end{equation}
Up to this point, the full action in the vortex sector reads
\begin{equation}
    S=S_{\rm vf}+S_J\;.\label{ym3}
\end{equation}
Note that the discussion can be trivially extended to a sector labeled by vortices carrying a distribution of weights $\beta^1,\dots, \beta^n$, i.e. $S_0=e^{i\chi^1\beta^1\cdot T}\dots e^{i\chi^n \beta^n\cdot T}$, where $\chi^i$ is multivalued when going around $\Omega_i$, and each $\beta^i$ takes values among the $N$ different fundamental weights. In this case, the physical values of the source would be
\begin{eqnarray}
    J^{\alpha}\Big|_{\rm phys}&=&J^{\bar{\alpha}}\Big|_{\rm phys}\;=\;0\;,\nonumber\\ J^q\Big|_{\rm phys}&=&i\sum\limits_{i=1}^n\beta_q^i\int d\sigma_1^id\sigma^i_2\sqrt{g(\sigma^i_1,\sigma^i_2)}\delta(x-x(\sigma^i_1,\sigma^i_2))\;,
\end{eqnarray}
where $x(\sigma^i_1,\sigma^i_2)$ is a parametrization of $\Omega_i$.

\section{Algebraic analysis of renormalizability}

In this section we analyze the center vortex sectors by employing the algebraic renormalization technique \cite{Piguet:1995er}. We will prove it at first order. Nevertheless, since this technique is recursive, the proof is valid to all orders in perturbation theory. 

\subsection{Ward Identities}\label{WARD}

As discussed in \cite{Fiorentini:2020tcn}, the action \eqref{ym2} for the vortex-free sector displays a rich set of Ward identities. It turns out that the same set of Ward identities can be accommodated for the action \eqref{ym3} of the center vortex sectors. For this aim, we have to include an additional external source term
\begin{eqnarray}
    S_{\rm ext}&=&s\int_x\left(K^a_\mu A^a_\mu+\bar{C}^ac^a+\bar{L}_I^ac^a_I+Q^a_I\zeta^a_I+B^a_Ib^a_I+\bar{L}^a_I\bar{c}^a_I +m^{ab}_{IJ}\zeta_I^a\xi_J^b+M^{ab}_I\bar{c}^a\zeta_I^b\right)\;,\nonumber\\
    &=&\int_x\left\{\frac{i}{g} K^a_\mu(D_\mu^{ab}c^b)
-\frac{1}{2}i\bar{C}^a f^{abc}c^{b}c^{c} 
-if^{abc}\bar{L}_I^ac_I^bc^c+Q_I^a (if^{abc}\zeta_I^bc^c+c_I^a)+if^{abc} B_I^a b_I^bc^c+\right.\nonumber\\
&-& \left. L_I^a (if^{abc}\bar{c}_I^bc^c+b_I^a)
+N^{ab}_I \bar{c}^a \zeta^b_I - M^{ab}_I b^a \zeta^b_I - M^{ab}_I\bar{c}^a(if^{bmn}\zeta_I^mc^n+c_I^b)+\right.\nonumber\\
&-&\left.n^{ab}_{IJ}\zeta_I^a\xi_J^b + m^{ab}_{IJ}\left[\lambda_I^a\zeta_J^b-\xi^a_I\left(if^{bcd}\zeta^c_Ic^d+c^b_I\right)\right]\right\}\;,\label{sext}
\end{eqnarray}
with $K$, $\bar{C}$, $\bar{L}$, $Q$, $B$, and $L$ being BRST invariant, while 
\begin{eqnarray}
sM^{ab}_I&=&N_I^{ab}\;,\nonumber\\
sN_I^{ab} &=& 0\;,\nonumber\\
    sm^{ab}_{IJ}&=&-n^{ab}_{IJ}\;,\nonumber\\ sn^{ab}_{IJ}&=&0\;.\label{brst6}
\end{eqnarray}
The large amount of sources introduced is necessary to control non linear symmetries as well as to ensure important symmetries to establish the renormalization of the theory. The quantum numbers of all sources are displayed in Table \ref{table3}. The final full action in the vortex sectors is then
\begin{equation}
    \Sigma=S+S_{ext}\;.\label{ym4}
\end{equation}
\begin{table}[h]
\centering
\begin{tabular}{|c|c|c|c|c|c|c|c|c|c|c|c|c|c|c|}
\hline
Sources &$K_\mu$&$\bar{C}$&$\bar{L}_I$&$L_I$&$B_I$&$Q_I$&$M_I$&$N_I$&$m_{IJ}$&$n_{IJ}$&$J$\\
\hline\hline
Mass dimension &3&4&3&3&3&3&1&1&2&2&2\\
\hline
Ghost number &$-1$&$-2$&$-2$&$0$&$-1$&$-1$&$0$&$1$&$0$&$1$&$0$\\
\hline
$\mathcal{Q}$-charge &0&0&$-1$&$1$&$1$&$-1$&$-1$&$-1$&$0$&$0$&$0$\\
\hline
Nature &F&B&B&B&F&F&B&F&B&F&B\\
\hline
\end{tabular}
\caption{Quantum numbers of the sources.}
\label{table3}
\end{table}
This action enjoys the following set of Ward identities:
\begin{itemize}
\item{The Slavnov-Taylor identity,
\begin{eqnarray}
\nonumber S(\Sigma) &=& \int_x \left(
\frac{\delta\Sigma}{\delta K^a_\mu}\frac{\delta\Sigma}{A^a_\mu} 
+  \frac{\delta\Sigma}{\delta\bar{L}^a_I}\frac{\delta\Sigma}{\delta c_I^a}  
+ \frac{\delta\Sigma}{\delta L^a_I} \frac{\delta\Sigma}{\delta \bar{c}_I^a} 
 +\frac{\delta\Sigma}{\delta B^a_I}\frac{\delta\Sigma}{\delta b_I} +\frac{\delta\Sigma}{\delta Q^a_I}\frac{\delta\Sigma}{\delta \zeta_I^a} 
 \right.\nonumber\\&+&\left. -
   b^a\frac{\delta\Sigma}{\delta\bar{c}^a}+\frac{\delta\Sigma}{\delta\bar{C}^a}\frac{\delta \Sigma}{\delta c^a}
  +N_I^{ab}\frac{\delta\Sigma}{\delta M_I^{ab}}
 +\lambda_I^a\frac{\delta \Sigma}{\delta\xi_I^a}-n_{IJ}^{ab}\frac{\delta\Sigma}{\delta m_{IJ}^{ab}}\right)
 \nonumber\\&+&U^2\frac{\delta \Sigma}{\delta\mu^2} 
 +\mathcal{K}\frac{\delta \Sigma}{\delta\kappa} 
 +\Lambda\frac{\delta \Sigma}{\delta\lambda}=0\;.
\end{eqnarray}
}

\item {The gauge-fixing equation,

\begin{equation}
\frac{\delta\Sigma}{\delta b^a}=if^{abc}\eta^b_I\zeta^c_I  -M^{ab}_I\zeta_I^b\;,
\label{WI_gfEq}
\end{equation}}

\item{The anti-ghost equation, 

\begin{equation}
\bar{\mathcal{G}}^a\Sigma=\left(\frac{\delta}{\delta\bar{c}^a}+M_I^{ab}\frac{\delta}{\delta Q_I^b}-if^{abc}\eta_I^b\frac{\delta}{\delta Q_I^c}\right)\Sigma
=N_{I}^{ab}\zeta_I^b \;.
\end{equation}
}

\item{The ghost number equation,

\begin{eqnarray}
\mathcal{N}_{\rm gh}\Sigma&=&\int d^{4}x\,\bigg(
c^{a}_I\frac{\delta}{\delta c^{a}_I} -\bar{c}^{a}_I\frac{\delta}{\delta \bar{c}^{a}_I}
+c^{a}\frac{\delta}{\delta c^{a}} -\bar{c}^{a}\frac{\delta}{\delta \bar{c}^{a}}
+U^{2}\frac{\delta}{\delta U^{2}}
+\mathcal{K}\frac{\delta}{\delta \mathcal{K}}
+\Lambda\frac{\delta}{\delta \Lambda}+
\nonumber\\
&-&
K^{a}\frac{\delta}{\delta K^a}-2\bar{C}^{a}\frac{\delta}{\delta \bar{C}^a}
-2\bar{L}^{a}_I\frac{\delta}{\delta \bar{L}^{a}_I}
-Q^{a}_I\frac{\delta}{\delta Q^{a}_I}-B^{a}_I\frac{\delta}{\delta B^{a}_I}
 +N_I^{ab}\frac{\delta}{\delta N_I^{ab}}+\nonumber\\&-&\xi_I^a\frac{\delta}{\delta \xi_I^a}+m_{IJ}^{ab}\frac{\delta}{\delta m_{IJ}^{ab}}
\bigg)\Sigma^{S_0}=0  \;.
\end{eqnarray}
}

\item{The global flavor symmetry,

\begin{eqnarray}
\mathcal{Q}\Sigma &=& \Bigg(\zeta_I^a\frac{\delta}{\delta \zeta_I^a}-b_I^a\frac{\delta }{\delta b_I^a} 
-\bar{c}_I^a\frac{\delta }{\delta  \bar{c}_I^a}
+c_I^a\frac{\delta }{\delta c_I^a}
-u^a_I\frac{\delta }{\delta u^a_I}
-Q_I^a\frac{\delta}{\delta Q_I^a}
+B_I^a\frac{\delta}{\delta B_I^a}
+L^a_I\frac{\delta}{\delta L^a_I}+
 \nonumber\\
 &-&
 \bar{L}^a_I\frac{\delta}{\delta \bar{L}^a_I}
-\kappa\frac{\delta }{\delta \kappa}  
 - 2\lambda\frac{\delta }{\delta \lambda}
 - \mathcal{K}\frac{\delta }{\delta \mathcal{K}}
 - 2\Lambda\frac{\delta }{\delta \Lambda}
  - N_I^{ab}\frac{\delta}{\delta N_I^{ab}} - M_I^{ab}\frac{\delta}{\delta M_I^{ab}}+\nonumber\\&-&\xi_I^a\frac{\delta}{\delta \xi_I^a}-\lambda_I^a\frac{\delta}{\delta \lambda_I^a}\Bigg)\Sigma=0 \;. 
\end{eqnarray}
 }
 
\item{The linearly broken rigid symmetry,

\begin{eqnarray}
\mathcal{R} \Sigma &=&\Bigg(\bar{c}_I^a\frac{\delta }{\delta b_I^a} 
+ \zeta_I^a\frac{\delta }{\delta c_I^a}-if^{abc}\eta_I^{a}\frac{\delta}{\delta N_I^{bc}}
-B_I^a\frac{\delta }{\delta \bar{L}_I^a}  + \bar{L}_I^a\frac{\delta }{\delta Q_I^a} 
-\kappa\frac{\delta }{\delta \mathcal{K}}-2\lambda\frac{\delta }{\delta\Lambda}-M_I^{ab}\frac{\delta}{\delta N_I^{ab}}+\nonumber\\&-&\xi_I^a\frac{\delta}{\delta \lambda_I^a}\Bigg)\Sigma= \bar{L}_I^ac_I^a +L_I^a\bar{c}_I^a  - \zeta_I^aQ_I^a\;.
\end{eqnarray}
}

\item{The ghost equation,}

\begin{eqnarray}
\mathcal{G}^a\Sigma&=&\Bigg(\frac{\delta}{\delta c^a}+(f^{abc}f^{cnm}\eta_I^n+if^{abn}M_I^{mn})\frac{\delta}{\delta N_I^{mb}}+i(f^{blc}f^{cma}J^l+im^{db}_{IJ}f^{dma})\frac{\delta}{\delta n_{IJ}^{mb}}\Bigg)\Sigma\nonumber\\
&=&if^{abc}\left( \bar{C}^{b}c^c + Q^{b}_I \zeta_I^c + \bar{L}^{b}_I c^c_I + L^{b}_I \bar{c}^c_I+ B^{b}_I b^c_I \right) 
+\frac{i}{g}D_\mu^{ab}K_\mu^{b}\;.
\end{eqnarray}

\item{The $J$ equation\footnote{ Notice that, due to this Ward identity, the variables $(J,m_{IJ})$ can enter the counterterm only through the combination $\delta_{IJ}f^{abc}J^a -m_{IJ}^{bc} $.},
}
\begin{equation}
\mathcal{J}^a\Sigma=\frac{\delta\Sigma}{\delta J^a}- f^{abc}\delta_{IJ}\frac{\delta\Sigma}{\delta m^{bc}_{IJ}}=0\;.
\label{WI_J}
\end{equation}

\item{Global symmetry in the boundary-conditions sector,}
\begin{equation}
\mathcal{F}\Sigma=\lambda^a_I\frac{\delta\Sigma}{\delta \lambda^a_I}+\xi^a_I\frac{\delta\Sigma}{\delta \xi^a_I}
-J^a\frac{\delta\Sigma}{\delta J^a}-n^{ab}_{IJ}\frac{\delta\Sigma}{\delta n^{ab}_{IJ}}
-m^{ab}_{IJ}\frac{\delta\Sigma}{\delta m^{ab}_{IJ}}=0\;.
\label{WI_FCh}
\end{equation}

\item The linearly broken $\lambda$ equation,
\begin{equation}
    \Lambda^a_I\Sigma=\frac{\delta\Sigma}{\delta\lambda^a_I}=f^{abc}\zeta^b_IJ^c\;.
\end{equation}

\end{itemize}

\subsection{The most general counterterm}

With the full action \eqref{ym4} at hand, we are now able to construct the most general counterterm $\Sigma_C$ compatible with all Ward identities in Section \ref{WARD}. Hence, we write the perturbative expansion of the quantum action $\Gamma$ at first order,
\begin{equation}
    \Gamma^{(1)}=\Sigma+\epsilon\Sigma_C\;,\label{qaction1}
\end{equation}
and impose on it all the Ward identities respected by the classical action $\Sigma$. A straightforward calculation leads to the following constraints for the counterterm,
\begin{eqnarray}
    \mathcal{B}_\Sigma\Sigma_C&=&0\;,\nonumber\\
    \frac{\delta\Sigma_C}{\delta b^a}&=&0\;,\nonumber\\
    \bar{\mathcal{G}}^a\Sigma_C&=&0\;,\nonumber\\
    \mathcal{N}_{gh}\Sigma_C&=&0\;,\nonumber\\
    \mathcal{Q}\Sigma_C&=&0\;,\nonumber\\
    \mathcal{R}\Sigma_C&=&0\;,\nonumber\\
    \mathcal{G}^a\Sigma_C&=&0\;,\nonumber\\
    \mathcal{J}^a\Sigma_C&=&0\;,\nonumber\\
    \mathcal{F}\Sigma_C&=&0\;,\nonumber\\
    \Lambda^a_I\Sigma_C&=&0\;.\label{count1}
\end{eqnarray}
Here,
\begin{eqnarray}
\mathcal{B}_{\Sigma} &=& \int_x \left(
\frac{\delta\Sigma}{\delta K^a_\mu}\frac{\delta}{A^a_\mu} +  \frac{\delta\Sigma}{\delta A^a_\mu}\frac{\delta}{\delta K^a_\mu} 
 + \frac{\delta\Sigma}{\delta L^a_I} \frac{\delta}{\delta \bar{c}_I^a} 
 +  \frac{\delta\Sigma}{\delta \bar{c}^a_I}\frac{\delta}{\delta L^a_I} +\frac{\delta\Sigma}{\delta\bar{L}^a_I}\frac{\delta}{\delta c_I^a} +  \frac{\delta\Sigma}{\delta c^a_I}\frac{\delta}{\delta \bar{L}^a_I} +
\right.\nonumber\\
&+& 
\frac{\delta\Sigma}{\delta B^a_I}\frac{\delta}{\delta b_I} +  \frac{\delta\Sigma}{\delta b^a_I}\frac{\delta}{\delta B^a_I}  + \frac{\delta\Sigma}{\delta Q^a_I}\frac{\delta}{\delta \zeta_I^a} 
 +  \frac{\delta\Sigma}{\delta \zeta^a_I}\frac{\delta}{\delta Q^a_I}+ \frac{\delta\Sigma}{\delta\bar{C}^a}\frac{\delta}{\delta c^a}  
+ \frac{\delta\Sigma}{\delta c^a}  \frac{\delta}{\delta\bar{C}^a}+
\nonumber\\
&+&N_I^{ab}\frac{\delta}{\delta M_I^{ab}}- \left.
b^a\frac{\delta}{\delta\bar{c}^a}
-n_{IJ}^{ab}\frac{\delta}{\delta m_{IJ}^{ab}}+\lambda_I^a\frac{\delta}{\delta\xi_I^a}\right)
 +U^2\frac{\delta }{\delta\mu^2} 
 +\mathcal{K}\frac{\delta }{\delta\kappa} 
 +\Lambda\frac{\delta }{\delta\lambda}\label{linST} 
\end{eqnarray}
is the linearized Slavnov-Taylor operator, which turns out to be nilpotent. Thence, the first equation in \eqref{count1} defines a cohomology problem for $\mathcal{B}_\Sigma$. The solution reads \cite{Piguet:1995er}
\begin{equation}
    \Sigma_C=\Delta_0+\mathcal{B}_\Sigma\Delta^{-1}\;,\label{count2}
\end{equation}
where $\Delta_0$ is the nontrivial part of the cohomology and $\mathcal{B}_\Sigma\Delta^{-1}$ is the trivial one. The non-trivial part is an integrated functional, polynomial in the fields, sources and their derivatives, with dimension 4, and vanishing ghost number. The quantity $\Delta^{-1}$ is also an integrated functional, polynomial in the fields, sources and their derivatives, with dimension 4, but with ghost number $-1$. Due to the rich set of constraints \eqref{count1}, it is a straightforward exercise to show that the nontrivial cohomology is the usual one in YM theory, namely
\begin{equation}
    \Delta_0=a_0S_{\rm YM}\;,
\end{equation}
with $a_0$ being an independent renormalization constant. For the trivial sector of the cohomology, we can write
\begin{eqnarray}
\Delta^{-1}=\bar{\Delta}^{-1}(\varphi)+D^{-1}(\varphi,\phi)\;,
\end{eqnarray}
where $\phi\equiv\{J,\lambda_I,\xi_I,m_{IJ},n_{IJ}\}$ and $\varphi$ stands for all the other fields, sources and parameters. This decomposition is a direct consequence of the Ward identity \eqref{WI_FCh} together with the quantum numbers of the sources involved. Then, it follows that $\bar{\Delta}^{-1}(\varphi)$ is identical to the full $\Delta^{-1}$ of the vortex-free sector obtained in Ref. \cite{Fiorentini:2020tcn}. Remarkably, after applying all remaining constraints in \eqref{count1}, one finds
\begin{eqnarray}
    \bar{\Delta}^{-1}&=&\int_x\Big[a_1(\bar{c}_I^a\partial^2\zeta_I^a+gf^{abc}\partial_\mu A_\mu^a\bar{c}_I^b\zeta_I^c+g^2f^{acm}f^{dbm}A_\mu^aA_\mu^b\bar{c}_I^c\zeta_I^d)+a_2f^{ IJK}f_{abc}\kappa\bar{c}_I^a\zeta_J^b\zeta_K^c+\nonumber\\& &+a_{3,IJKL}^{abcd}\lambda\bar{c}_I^a\zeta_J^b\zeta_K^c\zeta_L^d+a_{4}\mu^2\bar{c}_I^a\zeta_I^a\Big]\\
    D^{-1}&=&0\;,\label{deltas}
\end{eqnarray}
with $a_i$ being independent renormalization parameters. The tensor $a_{3,IJKL}^{abcd}$ has the same structure of  $\gamma_{IJKL}^{abcd}$. Therefore,
\begin{eqnarray}
\Sigma_C=\Sigma_C(\varphi)\;,
\end{eqnarray}
where $\Sigma_C(\varphi)$, given by
\begin{eqnarray}
    \Sigma_C(\varphi)&=&\int_x\left[\frac{a_0}{2}(\partial_\mu A^a_\nu)^2-\frac{a_0}{2}\partial_\nu A^a_\mu \partial_\mu A^a_\nu+\frac{a_0}{2}gf^{abc}A_\mu^aA_\nu^b\partial_\mu A_\nu^c+\frac{a_0}{4}g^2 f^{abc} f^{cde} A_\mu^a A_\nu^b A_\mu^d A_\nu^e+\right.\nonumber\\
    &+&\left.a_1(
    \partial_\mu b_I^a \partial_\mu \zeta_I^a +gf^{abc}\partial_\mu b_I^a A_\mu^b \zeta_I^c+gf^{abc}b_I^a\partial_\mu \zeta_I^b A_\mu^c +g^2 f^{abe}f^{cde}A_\mu^a b_I^bA_\mu^c \zeta_I^d+\right.\nonumber\\&+&\partial_\mu \bar{c}_I^a \partial_\mu c_I^a +gf^{abc}\partial_\mu \bar{c}_I^a A_\mu^b c_I^c +gf^{abc}\bar{c}_I^a\partial_\mu c_I^b A_\mu^c+g^2 f^{abe}f^{cde}A_\mu^a \bar{c}_I^bA_\mu^c c_I^d)\nonumber\\
    &+&\left.
a_2f^{IJK}f^{abc}(\mathcal{K}\bar{c}^a_I  \zeta_J^b \zeta_K^c-\kappa b_I^a\zeta_J^b\zeta_K^c-2\kappa\bar{c}_I^ac_J^b\zeta_K^c)+\right.\nonumber\\
&+&\left.
 a_{3,IJKL}^{abcd}(\Lambda \bar{c}^a_I  \zeta_J^b \zeta_K^c\zeta_L^d-\lambda b_I^a\zeta_J^b\zeta_K^c\zeta_L^d-3\lambda \bar{c}_I^ac_J^b\zeta_K^c\zeta_L^d)+\right.\nonumber\\
 &+&\left.
 a_4(U^2 \bar{c}_I^a\zeta_I^a-\mu^2 b_I^a\zeta_I^a-\mu^2\bar{c}_I^ac_I^a)\right] \;,
  \label{Final_CT}
\end{eqnarray}
is the vortex-free counterterm found in \cite{Fiorentini:2020tcn}.

\subsection{Quantum stability}

To prove stability, one has to show that the counterterm \eqref{Final_CT} can be absorbed in the original action \eqref{ym4} by means of a multiplicative redefinition of the fields, sources and parameters, \emph{i.e.},
\begin{equation}
    \Sigma(\Phi,\mathcal{S},P)+\epsilon\Sigma_C(\Phi,\mathcal{S},P)=\Sigma(\Phi_0,\mathcal{S}_0,P_0)\;,
\end{equation}
where $\Phi$ stands for the fields, $\mathcal{S}$ collects the sources, and $P$ contains the parameters. The bare fields are defined by the multiplicative renormalization
\begin{eqnarray}
    \Phi_0&=&\left(1+\frac{\epsilon}{2}z_\Phi\right)\Phi\;,\nonumber\\
    \mathcal{S}_0&=&\left(1+\epsilon z_{\mathcal{S}}\right)\mathcal{S}\;,\nonumber\\
    P_0&=&\left(1+\epsilon z_P\right)P\;.\label{bare}
\end{eqnarray}
As proven in \cite{Fiorentini:2020tcn}, the $\Sigma(\varphi)$ part is stable, and the factors $z_\varphi$ are the same as those of the vortex-free sector. Specifically,
\begin{align}
        &z_A=a_0\;, & z_g=-\frac{a_0}{2}\;,\nonumber \\
        &z_{c_I}=0\;, & z_{\bar{c}_I}=2 a_1\;,\nonumber \\
        &z_{\zeta_I}=0\;, & z_{b_I}=2 a_1\;,\nonumber \\
        &z_{\kappa}=-a_1-a_2\;,&z_{\mathcal{K}}=-a_1-a_2\;,\nonumber \\
         & z_\lambda= -a_1-a_4\;,&z_\Lambda= -a_1-a_4\;,\nonumber\\
        & z_{\mu^2}=-a_1-a_3\;,&z_{U^2}=-a_1-a_3\;, \nonumber\\
        &z_c=0\;, & z_{\bar{c}}=0\;, \nonumber\\
         & z_{\bar{C}}=0\;,&z_b=0\;, \nonumber\\
        &z_L=-a_1\;,&z_{\bar{L}}=0\;, \nonumber\\ 
        & z_K=-\frac{a_0}{2}\;, & z_B=-a_1\;,\nonumber\\ 
        & z_Q=0\;,
        &z_M=0\;,\nonumber \\
        & z_N =0 \; .\label{z}
    \end{align}
Moreover,
\begin{equation}
    z_n=z_m=z_J=-\frac{z_{\lambda_I}}{2}=-\frac{z_{\xi_I}}{2}\;.
\end{equation}
As there is no counterterm containing $J$, and $c_I$ and $\zeta_I$ do not renormalize, it is safe to set $z_n=z_m=z_J=z_{\lambda_I}=z_{\xi_I}=0$. Therefore, since the algebraic technique is recursive, the renormalizability of the model at all orders in perturbation theory is proven. The number of independent renormalizations is given by the number of independent renormalization parameters $a_i$, namely, five.

\section{Conclusions}
The search for a well defined quantization procedure for the Yang-Mills theory in the nonperturbative regime has  attracted a lot of activity for many years. Many proposals have been analyzed, always considering gauge fixing procedures which are global in configuration space. Global conditions lead to the Gribov problem, which has been tackled by restricting the configuration space to be path-integrated. A different  way out  was raised at the end of Ref. \cite{Singer:1978dk}, where a superposition of infinitely many local  gauge-fixings was proposed. Recently, a particular realization of this general scenario was implemented by means of a partition of the configuration space into sectors labeled by topological degrees of freedom. In this work, we showed for the first time that this path is in principle calculable. Namely, we established the all-orders perturbative renormalizability of the procedure in sectors labeled by oriented center vortices. Remarkably, as the counterterms are the same as those of the vortex-free sector, no new parameters had to be introduced. In a future work, it would be important to explicitly calculate an approximation to the partial contributions defined in Eq. \eqref{general-1}. At large distances, they are expected to contain terms proportional to the area and to the square of the extrinsic curvature of $\Omega$,  the closed worldsurface where the center-vortex guiding centers are located.  This points to the idea that Singer's no go theorem is the fundamental reason behind a first-principles YM center-vortex ensemble. Furthermore, this could establish a connection with phenomenological ensembles of center vortices, which are known to successfully reproduce the properties of the confining string (see the reviews \cite{greensite_book,universe7080253}, and references therein).

\section*{Acknowledgments}

The Coordena\c c\~ao de Aperfei\c coamento de Pessoal de N\'{\i}vel Superior - Brasil (CAPES) – Finance Code 001, the Deutscher Akademischer Austauschdienst (DAAD) and the Conselho Nacional de Desenvolvimento Cient\'{\i}fico e Tecnol\'{o}gico  (CNPQ) are acknowledged for the financial support. 

\appendix

\section{Lie Algebra conventions}
\label{Ap1}

In this work, following Ref. \cite{Oxman:2012ej} and references therein, we use a basis for $\mathfrak{su}(N)$, the Lie Algebra of SU(N), which relies on the Cartan decomposition. The first $N-1$ elements are given by the generators $T_q$, $q=1,\dots,N-1$ of the Cartan subalgebra, also known as the maximal torus of $\mathfrak{su}(N)$, since all of its elements commute with each other:
\begin{align}
    [T_q,T_p]=0\;.
\end{align}
Then, we define the eigenvectors $E_{\alpha}$ of the adjoint action of $T_q$:
\begin{align}
    [T_q,E_\alpha]=\alpha_q E_\alpha\;. \label{adjactio}
\end{align}
The eigenvalues $\alpha_q$, $q=1,\dots,N-1$ are known as the roots of $\mathfrak{su}(N)$. It is possible to define a notion of ordering of these objects, where a root is said to be positive if and only if its last nonvanishing component is positive. The number of negative and positive roots is the same. This follows from the fact that if $\alpha$ is a root, then $-\alpha$ is also a root, with $E_{-\alpha}=E_\alpha^\dagger$. This may be obtained by taking the Hermitian conjugate of Eq. \eqref{adjactio}. Then, the remaining $N(N-1)$ Hermitian generators are defined as
\begin{align}
   & T_\alpha=\frac{E_{\alpha}+E^\dagger_\alpha}{\sqrt{2}} \\&
   T_{\bar{\alpha}}=\frac{E_{\alpha}-E^\dagger_\alpha}{i\sqrt{2}}\;.
\end{align}
 We denote the elements of the basis $(T_q,T_\alpha,T_{\bar{\alpha}})$ collectively by $T^a$, always with a latin index different than $p,q$, which we use only for the Cartan generators. The commutation relations of this basis which are relevant for the purposes of this work are 
\begin{align}
    &[T_q,T_p]=0\;,\\& [T_q,T_\alpha]= i\alpha_q T_{\bar{\alpha}}\;, \\&
    [T_q,T_{\bar{\alpha}}]=-i\alpha_q T_\alpha\;.
\end{align}
These relations, together with the fact that the commutators between root generators never involve Cartan generators, imply that $f^{qbc}$ is nonvanishing only when $b=\alpha$ and $c=\bar{\alpha}$, or $b=\bar{\alpha}$ and $c=\alpha$. Finally, we remark that this basis is orthonormal with respect to the Killing metric
\begin{align}
    (A,B)={\rm Tr}({\rm Ad}(A){\rm Ad}(B))\;, \label{killing}
\end{align}
where ${\rm Ad}()$ stands for the adjoint representation of the Lie Algebra.

\providecommand{\href}[2]{#2}\begingroup\raggedright\endgroup


\begin{thebibliography}{10}

\bibitem{Singer:1978dk}
I.~M. Singer, ``{Some Remarks on the Gribov Ambiguity}''.
\href{http://dx.doi.org/10.1007/BF01609471}{{\em Commun. Math. Phys.}
  {\bfseries 60} (1978) 7--12}.

\bibitem{Gribov:1977wm}
V.~N. Gribov, ``{Quantization of Nonabelian Gauge Theories}''.
\href{http://dx.doi.org/10.1016/0550-3213(78)90175-X}{{\em Nucl. Phys.}
  {\bfseries B139} (1978) 1}.

\bibitem{Sobreiro:2005ec}
R.~F. Sobreiro and S.~P. Sorella, ``{Introduction to the Gribov ambiguities in
  Euclidean Yang-Mills theories}''. in {\em {13th Jorge Andre Swieca Summer
  School on Particle and Fields Campos do Jordao, Brazil, January 9-22, 2005}}.
\newblock
2005.
\newblock

\bibitem{Zwanziger:1989mf}
D.~Zwanziger, ``{Local and Renormalizable Action From the Gribov Horizon}''.
\href{http://dx.doi.org/10.1016/0550-3213(89)90122-3}{{\em Nucl. Phys.}
  {\bfseries B323} (1989) 513--544}.

\bibitem{Zwanziger:1992qr}
D.~Zwanziger, ``{Renormalizability of the critical limit of lattice gauge
  theory by BRS invariance}''.
\href{http://dx.doi.org/10.1016/0550-3213(93)90506-K}{{\em Nucl. Phys.}
  {\bfseries B399} (1993) 477--513}.

\bibitem{Dudal:2005na}
D.~Dudal, R.~F. Sobreiro, S.~P. Sorella, and H.~Verschelde, ``{The Gribov
  parameter and the dimension two gluon condensate in Euclidean Yang-Mills
  theories in the Landau gauge}''.
\href{http://dx.doi.org/10.1103/PhysRevD.72.014016}{{\em Phys. Rev.} {\bfseries
  D72} (2005) 014016}.

\bibitem{Dudal:2008sp}
D.~Dudal, J.~A. Gracey, S.~P. Sorella, N.~Vandersickel, and H.~Verschelde, ``{A
  Refinement of the Gribov-Zwanziger approach in the Landau gauge: Infrared
  propagators in harmony with the lattice results}''.
\href{http://dx.doi.org/10.1103/PhysRevD.78.065047}{{\em Phys. Rev.} {\bfseries
  D78} (2008) 065047}.

\bibitem{Dudal:2011gd}
D.~Dudal, S.~P. Sorella, and N.~Vandersickel, ``{The dynamical origin of the
  refinement of the Gribov-Zwanziger theory}''.
\href{http://dx.doi.org/10.1103/PhysRevD.84.065039}{{\em Phys. Rev.} {\bfseries
  D84} (2011) 065039}.

\bibitem{Capri:2015ixa}
M.~A.~L. Capri, D.~Dudal, D.~Fiorentini, M.~S. Guimaraes, I.~F. Justo, A.~D.
  Pereira, B.~W. Mintz, L.~F. Palhares, R.~F. Sobreiro, and S.~P. Sorella,
  ``{Exact nilpotent nonperturbative BRST symmetry for the Gribov-Zwanziger
  action in the linear covariant gauge}''.
  \href{http://dx.doi.org/10.1103/PhysRevD.92.045039}{{\em Phys. Rev. D}
  {\bfseries 92} no.~4, (2015) 045039}.

\bibitem{Capri:2016aqq}
M.~A.~L. Capri, D.~Dudal, D.~Fiorentini, M.~S. Guimaraes, I.~F. Justo, A.~D.
  Pereira, B.~W. Mintz, L.~F. Palhares, R.~F. Sobreiro, and S.~P. Sorella,
  ``{Local and BRST-invariant Yang-Mills theory within the Gribov horizon}''.
\href{http://dx.doi.org/10.1103/PhysRevD.94.025035}{{\em Phys. Rev.} {\bfseries
  D94} no.~2, (2016) 025035}.

\bibitem{rudin}
W.~Rudin, {\em {Real and complex analysis}}.
\newblock New York: McGraw-Hill, 1987.

\bibitem{koba}
S.~Kobayashi and K.~Nomizu, {\em {Foundation of Differential Geometry, Vol. 1.
  }}.
\newblock Inderscience Publishers, New York, London, Sydney, 1987.

\bibitem{Oxman:2015ira}
L.~E. Oxman and G.~C. Santos-Rosa, ``{Detecting topological sectors in
  continuum Yang-Mills theory and the fate of BRST symmetry}''.
\href{http://dx.doi.org/10.1103/PhysRevD.92.125025}{{\em Phys. Rev. D}
  {\bfseries 92} (2015) 125025}.

\bibitem{PhysRevD.103.114010}
D.~Fiorentini, D.~R. Junior, L.~E. Oxman, G.~M. Sim\~oes, and R.~F. Sobreiro,
  ``{Study of Gribov copies in a Yang-Mills ensemble}''.
  \href{http://dx.doi.org/10.1103/PhysRevD.103.114010}{{\em Phys. Rev. D}
  {\bfseries 103} (2021) 114010}.

\bibitem{DelDebbio:1996lih}
L.~Del~Debbio, M.~Faber, J.~Greensite, and S.~Olejnik, ``{Center dominance and
  Z(2) vortices in SU(2) lattice gauge theory}''.
\href{http://dx.doi.org/10.1103/PhysRevD.55.2298}{{\em Phys. Rev. D} {\bfseries
  55} (1997) 2298--2306}.

\bibitem{Langfeld:1997jx}
K.~Langfeld, H.~Reinhardt, and O.~Tennert, ``{Confinement and scaling of the
  vortex vacuum of SU(2) lattice gauge theory}''.
\href{http://dx.doi.org/10.1016/S0370-2693(97)01435-4}{{\em Phys. Lett. B}
  {\bfseries 419} (1998) 317--321}.

\bibitem{DelDebbio:1998luz}
L.~Del~Debbio, M.~Faber, J.~Giedt, J.~Greensite, and S.~Olejnik, ``{Detection
  of center vortices in the lattice Yang-Mills vacuum}''.
\href{http://dx.doi.org/10.1103/PhysRevD.58.094501}{{\em Phys. Rev. D}
  {\bfseries 58} (1998) 094501}.

\bibitem{Faber:1997rp}
M.~Faber, J.~Greensite, and S.~Olejnik, ``{Casimir scaling from center
  vortices: Towards an understanding of the adjoint string tension}''.
\href{http://dx.doi.org/10.1103/PhysRevD.57.2603}{{\em Phys. Rev. D} {\bfseries
  57} (1998) 2603--2609}.

\bibitem{deForcrand:1999our}
P.~de~Forcrand and M.~D'Elia, ``{On the relevance of center vortices to QCD}''.
\href{http://dx.doi.org/10.1103/PhysRevLett.82.4582}{{\em Phys. Rev. Lett.}
  {\bfseries 82} (1999) 4582--4585}.

\bibitem{Ambjorn:1999ym}
J.~Ambjorn, J.~Giedt, and J.~Greensite, ``{Vortex structure versus monopole
  dominance in Abelian projected gauge theory}''.
\href{http://dx.doi.org/10.1088/1126-6708/2000/02/033}{{\em JHEP} {\bfseries
  02} (2000) 033}.

\bibitem{Engelhardt:1999fd}
M.~Engelhardt, K.~Langfeld, H.~Reinhardt, and O.~Tennert, ``{Deconfinement in
  SU(2) Yang-Mills theory as a center vortex percolation transition}''.
\href{http://dx.doi.org/10.1103/PhysRevD.61.054504}{{\em Phys. Rev. D}
  {\bfseries 61} (2000) 054504}.

\bibitem{Engelhardt:1999xw}
M.~Engelhardt and H.~Reinhardt, ``{Center projection vortices in continuum
  Yang-Mills theory}''.
\href{http://dx.doi.org/10.1016/S0550-3213(99)00727-0}{{\em Nucl. Phys. B}
  {\bfseries 567} (2000) 249}.

\bibitem{Bertle:2001xd}
R.~Bertle, M.~Engelhardt, and M.~Faber, ``{Topological susceptibility of
  Yang-Mills center projection vortices}''.
\href{http://dx.doi.org/10.1103/PhysRevD.64.074504}{{\em Phys. Rev. D}
  {\bfseries 64} (2001) 074504}.

\bibitem{Reinhardt:2001kf}
H.~Reinhardt, ``{Topology of center vortices}''.
\href{http://dx.doi.org/10.1016/S0550-3213(02)00130-X}{{\em Nucl. Phys. B}
  {\bfseries 628} (2002) 133--166}.

\bibitem{Gattnar:2004gx}
J.~Gattnar, C.~Gattringer, K.~Langfeld, H.~Reinhardt, A.~Schafer, S.~Solbrig,
  and T.~Tok, ``{Center vortices and Dirac eigenmodes in SU(2) lattice gauge
  theory}''.
\href{http://dx.doi.org/10.1016/j.nuclphysb.2005.03.027}{{\em Nucl. Phys. B}
  {\bfseries 716} (2005) 105--127}.

\bibitem{Fiorentini:2020tcn}
D.~Fiorentini, D.~R. Junior, L.~E. Oxman, and R.~F. Sobreiro,
  ``{Renormalizability of the center-vortex free sector of Yang-Mills
  theory}''. \href{http://dx.doi.org/10.1103/PhysRevD.101.085007}{{\em Phys.
  Rev. D} {\bfseries 101} no.~8, (2020) 085007}.

\bibitem{Piguet:1995er}
O.~Piguet and S.~P. Sorella, ``{Algebraic renormalization: Perturbative
  renormalization, symmetries and anomalies}''.
\href{http://dx.doi.org/10.1007/978-3-540-49192-7}{{\em Lect. Notes Phys.
  Monogr.} {\bfseries 28} (1995) 1--134}.

\bibitem{Daniel:1979ez}
M.~Daniel and C.~M. Viallet, ``{The Geometrical Setting of Gauge Theories of
  the {Yang-Mills} Type}''.
  \href{http://dx.doi.org/10.1103/RevModPhys.52.175}{{\em Rev. Mod. Phys.}
  {\bfseries 52} (1980) 175}.

\bibitem{Nakahara:2003nw}
M.~Nakahara, {\em {Geometry, topology and physics}}.
\newblock Boca Raton, USA: Taylor \& Francis (2003) 573 p, 2003.

\bibitem{Bertlmann:1996xk}
R.~A. Bertlmann, {\em {Anomalies in quantum field theory}}.
\newblock Oxford, UK: Clarendon (1996) 566 p. (International series of
  monographs on physics: 91), 1996.

\bibitem{Piguet:1984js}
O.~Piguet and K.~S., ``{Gauge Independence in Ordinary {Yang-Mills}
  Theories}''.
\href{http://dx.doi.org/10.1016/0550-3213(85)90545-0}{{\em Nucl. Phys.}
  {\bfseries B253} (1985) 517--540}.

\bibitem{Golestanian:1998bx}
R.~Golestanian and M.~Kardar, ``{Path integral approach to the dynamic Casimir
  effect with fluctuating boundaries}''.
  \href{http://dx.doi.org/10.1103/PhysRevA.58.1713}{{\em Phys. Rev. A}
  {\bfseries 58} (1998) 1713--1722}.

\bibitem{Symanzik:1969ek}
K.~Symanzik, ``{Renormalizable models with simple symmetry breaking. 1.
  Symmetry breaking by a source term}''.
  \href{http://dx.doi.org/10.1007/BF01645494}{{\em Commun. Math. Phys.}
  {\bfseries 16} (1970) 48--80}.

\bibitem{Moss:1996ip}
I.~G. Moss and P.~J. Silva, ``{BRST invariant boundary conditions for gauge
  theories}''. \href{http://dx.doi.org/10.1103/PhysRevD.55.1072}{{\em Phys.
  Rev. D} {\bfseries 55} (1997) 1072--1078}.

\bibitem{greensite_book}
J.~Greensite, {\em {An Introduction to the Confinement Problem}}.
\newblock Springer Nature, Switzerland, 2020.

\bibitem{universe7080253}
D.~R. Junior, L.~E. Oxman, and G.~M. Simões, ``From center-vortex ensembles to
  the confining flux tube''.
  \href{http://dx.doi.org/10.3390/universe7080253}{{\em Universe} {\bfseries 7}
  no.~8, (2021) }.

\bibitem{Oxman:2012ej}
L.~E. Oxman, ``{Confinement of quarks and valence gluons in SU(N)
  Yang-Mills-Higgs models}''.
\href{http://dx.doi.org/10.1007/JHEP03(2013)038}{{\em JHEP} {\bfseries 03}
  (2013) 038}.

\end{thebibliography}
\end{document}